\documentclass[a4paper,fleqn,usenatbib,useAMS]{mnras}

\usepackage{newtxtext,newtxmath}

\usepackage[T1]{fontenc}
\usepackage{ae,aecompl}
\usepackage{hyperref}

\usepackage{natbib}
\bibpunct{(}{)}{;}{a}{,}{,}

\usepackage{graphicx}	
\usepackage{amsmath}	
\usepackage{amssymb}	

\usepackage{threeparttable}
\usepackage{lipsum}
\usepackage{mathtools}
\usepackage{cuted}
\usepackage{float}

\title[ULXs as magnetized advective accretion flows]{Ultra-luminous X-ray sources as magnetically powered sub-Eddington advective accretion flows around stellar mass black holes}

\author[T. Mondal and B. Mukhopadhyay]{
Tushar Mondal\thanks{E-mail: mtushar@iisc.ac.in (TM)} and
Banibrata Mukhopadhyay\thanks{bm@iisc.ac.in (BM)}
\\
Department of Physics, Indian Institute of Science, Bangalore 560012, India
}

\date{Accepted 2018 August 24. Received 2018 August 15; in original form 2018 July 7}

\pubyear{2018}

\begin{document}
\label{firstpage}
\pagerange{\pageref{firstpage}--\pageref{lastpage}}
\maketitle

\begin{abstract}
In order to explain unusually high luminosity and spectral nature of ultra-luminous X-ray sources 
(ULXs), some of the
underlying black holes are argued to be of intermediate mass, between several 
tens to million solar masses. Indeed, there is a long standing 
question of missing mass of intermediate range of black holes. However, as
some ULXs are argued to be neutron stars too, often their unusual high luminosity
is argued by super-Eddington accretions. Nevertheless, all the models
are based on non-magnetized or weakly magnetized accretion. There are, however, 
evidences that magnetic fields in accretion discs/flows around
a stellar mass black hole could be million Gauss. Such a magnetically
arrested accretion flow plausibly plays a key role to power many combined 
disc-jet/outflow systems. Here we show that flow energetics
of a 2.5-dimensional advective magnetized accretion disc/outflow system around a stellar mass black hole
are sufficient to explain power of ULXs in their hard states. Hence, they
are neither expected to have intermediate mass black holes 
nor super-Eddington accretors. We suggest that at least some 
ULXs are magnetically powered sub-Eddington accretors 
around a stellar mass black hole.
\end{abstract}




\section{INTRODUCTION}


Ultra-luminous X-ray sources (ULXs) are very bright, point-like, non-nuclear X-ray emitters found in nearby galaxies. Their apparent luminosities, assuming isotropic emission, are in the range of $3 \times 10^{39}-10^{41}\ ergs \ s^{-1}$, which exceed the Eddington luminosity limit of a neutron star or even that of the heaviest stellar-mass black hole $(\sim 20M_{\odot})$ \citep{2006ARA&A..44..323F}. Here, the Eddington limit is defined as
\begin{equation}
L_{Edd}=\frac{4\pi c G M m_{p}}{\sigma_{T}}\simeq 1.3\times 10^{38} \left(\frac{M}{M_{\odot}} \right) \text{erg s$^{-1}$},
\end{equation}
where $M$ is the mass of the accretor, $m_{p}$ the proton mass, $\sigma_{T}$ the Thomson scattering cross-section, $G$ the Newton's gravitation constant and 
$c$ the speed of light.

Three alternate physical scenarios have been proposed to explain the large apparent luminosities of ULXs. One possibility is that they might be powered by accretion on to intermediate-mass black holes (IMBHs) with masses in the range of $10^{2}-10^{4}\ M_{\odot}$. Second, they can be stellar-mass black holes, achieved super-Eddington luminosities through slim-disc model \citep{2003ApJ...597..780E} or radiation pressure dominated geometrically thin accretion disc model \citep{2002ApJ...568L..97B} as a result of the nonlinear development of ``photon-bubble instability" \citep{1998MNRAS.297..929G}. A third scenario is beamed emission from a stellar mass black hole system, either through relativistic boosting along our line of sight \citep{2002A&A...382L..13K} or through geometric beaming effect \citep{2001ApJ...552L.109K}. A combination of supper-Eddington and mild beamed emission from stellar mass black hole can also be a plausible mechanism to explain their large apparent luminosities \citep{2007MNRAS.377.1187P}.

Indeed in some rare cases, dodging of this Eddington limit is possible. In highly magnetized neutron stars, the presence of large magnetic fields $B\gtrsim 10^{12} \ G$ suppresses the electron scattering cross-section \citep{1979PhRvD..19.2868H} and, hence, reduces the effect of radiation pressure and increases the effective Eddington luminosity. 
In addition, the strong magnetic fields of the neutron star disrupt the accretion flow at the Alfven radius, and the matter is funneled along the field lines onto the magnetic poles. This geometry also provides apparent super-Eddington luminosity, as radiation can escape from the sides of the column \citep{1976MNRAS.175..395B}, perpendicular to the magnetic field in a ``fan beam" pattern.

The important evidence supporting IMBHs scenario is the presence of soft excesses in the energy spectra of some ULXs. X-ray spectra in a number of ULXs are shown to be well fitted with the  combined multicolour disc blackbody and power-law continuum model, similar to Galactic black hole binaries. The key difference is that the derived disc temperatures for ULX spectra are $0.1-0.3$ keV \citep{2004ApJ...614L.117M}, much lower than that for stellar mass black holes in their high state (at around $1$ keV). 
The cool accretion disc suggests a missing population of high-state of IMBHs. However, this cool accretion disc model has been disputed extensively. \cite{2006MNRAS.371..673G} argued that the soft excess could be a soft deficit depending on the energy range over which the power-law continuum is modeled. They showed that the spectra could be fitted equally well with a combination of smeared emission and absorption lines from highly ionized, fast outflow surrounding the primary X-ray source. Hence, they suggested that those components should not be taken as evidence for accretion disc emission, nor provided reliable measure of black hole masses. \cite{2006MNRAS.371.1216D} explained this cool disc by ``disc-corona coupling" model, where the optically thick Comptonizing corona over the inner disc drains power from the hot disc material.

In the context of searching for the true physical nature of ULXs, one or two ULXs might be intermediate mass black holes \citep{2009Natur.460...73F}. The more recent argument (see \citealt{2007MNRAS.377.1187P}; \citealt{2011NewAR..55..166F} for reviews, \citealt{2002ApJ...568L..97B}; \citealt{2014Natur.514..198M} for theoretical disputes) is that the majority of ULXs are stellar mass black holes. Recent identification of coherent pulsations in three sources (M82 X-2, \citealt{2014Natur.514..202B}; NGC 7793 P13, \citealt{2016ApJ...831L..14F}; and, NGC 5907 ULX-1, \citealt{2017Sci...355..817I}) has brought support to the perspective that some ULXs likely host a neutron star. Most ULXs with steep power law, soft excess and/or high energy downturn can well be explained by different models. Nevertheless the interpretation of a significant fraction of ULXs with a hard power-law spectrum remains mysterious. \cite{2011AN....332..330S} already pointed out regarding this long-standing issue (see also \citealt{2006ApJ...649..730W}).

In this letter, we propose a magnetized disc-outflow coupled model to address a plausible mechanism of finding the hidden nature of hard-state ULXs. The disc threaded by ordered magnetic fields provides the most efficient way of tapping the gravitational potential energy of black hole liberated through accretion to power jets/outflows \citep{1982MNRAS.199..883B}. As the pseudo-Newtonian framework considered here does not capture full general relativistic effect, the present model is inefficient of tapping the rotational energy of black hole \citep{1977MNRAS.179..433B}, unlike magnetically arrested disc (MAD) model \citep{2011MNRAS.418L..79T}. The magneto-centrifugally driven outflows are more plausible to emerge from the hot puffed up region of the advective accretion flow. Also, vertically inflated strong toroidal fields can enhance the outflow power in the form of ``magnetic tower" \citep{2004ApJ...605..307K}. We suggest that the observed hard-state ULXs are actually geometrically thick, highly magnetized, advective but sub-Eddington accretion flows orbiting stellar mass black holes and hence no need to incorporate the existence of the missing class of IMBHs, nor super-Eddington accretions.

The letter is organised as follows. In Section 2, we recall the spectral classifications of ULXs along with some hard-state sources, the heart of interest of this letter. In Section 3, we model the coupled disc-outflow symbiosis for magnetized advective accretion flows. Subsequently, we discuss our results, in particular focusing on the energetics of the accretion induced outflows, in Section 4. Finally we end with discussions and conclusions in Sections 5 and 6 respectively.  


\section{SPECTRAL CLASSIFICATIONS}

Since ULXs are believed to be powered by accretion on to black holes (in some rare cases neutron stars), a keen knowledge of the spectral properties of Galactic black hole binaries could be essential to interpret their peculiarities. Traditionally, Galactic black hole X-ray binaries pass through three most familiar canonical states \citep{2006ARA&A..44...49R, 2004MNRAS.355.1105F}: low/hard (LH), high/soft (HS) and very high (VH) states. 
The LH state is dominated generally by radiatively inefficient, quasi-spherical, sub-Keplerian, advective disc and/or jets at lower mass accretion rate and is well explained by a hard power-law component $(\text{the photon index} \ \Gamma \sim 1.4 \ \text{to} \ 1.8)$. Several ULXs with this hard power-law dominated state are listed in Table~\ref{tab:table1} with measurement in the $0.3-10$ keV energy band. However, unlike canonical galactic black hole sources, their luminosity is not low. Hence, they are not really in low/hard state \citep{2013MNRAS.435.1758S}. The hard spectrum has been thought to arise due 
to inverse-Compton scattering of soft photons from the accretion disc by either hot optically thin corona \citep{1977ApJ...218..247L}, or sub-Keplerian flow surrounding it producing hot shock close to the black hole \citep{1995ApJ...455..623C}, or due to synchrotron emission at the jet-footprint \citep{2005ApJ...635.1203M}. It also could be produced by advection-dominated accretion flow \cite[ADAF,][]{1995ApJ...452..710N}. The HS state is dominated by optically thick, geometrically thin, Keplerian accretion disc and is well explained by a multicolour disc blackbody, sometimes with a little contribution from hard tail. In VH state, the spectrum consists of both a disc component and an unbroken power law component extended to higher energies. The photon index is steeper $(\Gamma \gtrsim 2.5)$ than that found in a LH state.
\begin{table}
	\centering
	\caption{Some ULX sources in a hard power-law dominated state.}
	\label{tab:table1}
	\begin{threeparttable}
		\begin{tabular}{lccr} 
			\hline
			Source & $\Gamma$ & $L_{0.3-10 \ \text{keV}}$  & Ref.\\
			&          & $(10^{40}$ erg  s$^{-1} )$ & \\
			\hline       
			NGC 3628 X1 & $1.8^{+0.2}_{-0.2}$ & 1.1 & 1\\       
			\hline
			M99 X1 & $1.7^{+0.1}_{-0.1}$ & 1.9 & 2\\
			\hline
			Antennae X-11 & $1.76^{+0.05}_{-0.05}$ & 2.11 & 3\\
			& $1.68^{+0.06}_{-0.06}$ & 1.38 &  \\
			Antennae X-16 & $1.35^{+0.03}_{-0.04}$ & 1.82 &  \\
			& $1.2^{+0.14}_{-0.10}$ & 0.90 &  \\
			Antennae X-42 & $1.73^{+0.10}_{-0.11}$ & 0.96 &  \\
			& $1.66^{+0.05}_{-0.06}$ & 1.00 &  \\
			Antennae X-44 & $1.74^{+0.04}_{-0.04}$ & 1.28 &  \\
			& $1.63^{+0.09}_{-0.09}$ & 1.48 &  \\
			\hline
			Holmberg IX X-1 & $1.9^{+0.1}_{-0.02}$ & 1.0 & 4\\
			\hline
			NGC 1365 X1 & $1.74^{+0.12}_{-0.11}$ & 2.8 & 5\\
			& $1.80^{+0.04}_{-0.05}$ & 0.53 &  \\
			NGC 1365 X2 & $1.23^{+0.25}_{-0.19}$ & 3.7 &  \\
			& $1.13^{+0.09}_{-0.10}$ & 0.15 &  \\
			\hline
			M82 X42.3+59 & $1.44^{+0.09}_{-0.09}$ & 1.13 & 6\\
			& $1.33^{+0.13}_{-0.13}$ & 1.51 &  \\            
			\hline
		\end{tabular}
		\begin{tablenotes}
			\item References: $(1)$ \cite{2001ApJ...560..707S}; $(2)$ \cite{2006MNRAS.372.1531S}; $(3)$ \cite{2009ApJ...696.1712F}; $(4)$ \cite{2009ApJ...702.1679K}; $(5)$ \cite{2009ApJ...695.1614S}; $(6)$ \cite{2010ApJ...710L.137F}.
		\end{tablenotes}
	\end{threeparttable}
\end{table}

\section{MODELLING THE COUPLED DISC-OUTFLOW SYSTEM}

We propose a magnetized combined disc-outflow model. Unlike previous exploration \cite[e.g.][]{2004ApJ...616..669K}, here the dynamics is primarily controlled by large scale magnetic stress. In $1.5-$dimension, magnetized advective disc models were already proposed by us \citep{2015ApJ...807...43M, 2018MNRAS.476.2396M}, without any vertical flow. For the present purpose, $2.5-$dimensional description is necessary. In this $2.5-$dimensional disc-outflow symbiotic model, we describe magnetized, viscous, advective accretion flows around black holes, in the pseudo-Newtonian framework with \cite{2002ApJ...581..427M} potential. Here, we assume a steady and axisymmetric flow such that $\partial /\partial t \equiv \partial /\partial \phi \equiv 0$ and all the flow variables, namely, radial velocity $(v_{r})$, specific angular momentum $(\lambda)$, vertical or outflow velocity $(v_{z})$, mass density $(\rho)$, fluid pressure $(p)$, radial $(B_{r})$, azimuthal $(B_{\phi})$ and vertical $(B_{z})$ components of magnetic fields, as functions of both radial and vertical coordinates.

Throughout in our calculations, we express any length variable in units of $r_g=GM_{BH}/c^{2}$, where $G$ is the Newton's gravitational constant, $M_{BH}$ the mass of the black hole, and $c$ the speed of light. Accordingly, we also express the velocities in units of $c$ and the specific angular momentum in $GM_{BH}/c$ to make all the variables dimensionless. Hence, the equation of continuity, the momentum balance equations, the equation for no magnetic monopole, the magnetic induction equations and the energy equation are respectively,

\begin{equation}
\frac{1}{r} \frac{\partial}{\partial r}\left(r\rho v_{r}\right)+\frac{\partial}{\partial z}\left(\rho v_{z}\right)=0,
\end{equation}
\begin{multline}
v_{r}\frac{\partial v_{r}}{\partial r}+v_{z}\frac{\partial v_{r}}{\partial z}-\frac{\lambda^{2}}{r^{3}}+\frac{1}{\rho}\frac{\partial p}{\partial r}+F=\frac{1}{\rho}\frac{\partial W_{rz}}{\partial z}\\
+\frac{1}{4 \pi  \rho}\left[-\frac{B_{\phi}}{r}\frac{\partial}{\partial r}\left(rB_{\phi}\right)+B_{z}\left(\frac{\partial B_{r}}{\partial z}-\frac{\partial B_{z}}{\partial r}\right)\right],
\end{multline}
\begin{multline}
v_{r}\frac{\partial \lambda}{\partial r}+v_{z}\frac{\partial \lambda}{\partial z}=\frac{r}{\rho}\left[\frac{1}{r^{2}}\frac{\partial}{\partial r}\left(r^{2}W_{r\phi}\right)+\frac{\partial W_{\phi z}}{\partial z}\right]\\
+\frac{r}{4\pi \rho}\left[\frac{B_{r}}{r}\frac{\partial}{\partial r}\left(rB_{\phi}\right)+B_{z}\frac{\partial B_{\phi}}{\partial z}\right],
\end{multline}
\begin{multline}
v_{r}\frac{\partial v_{z}}{\partial r}+v_{z}\frac{\partial v_{z}}{\partial z}+\frac{1}{\rho}\frac{\partial p}{\partial z}+\frac{Fz}{r}=\frac{1}{r\rho}\frac{\partial}{\partial r}\left(rW_{rz}\right)\\
+\frac{1}{4 \pi  \rho}\left[B_{r}\left(\frac{\partial B_{z}}{\partial r}-\frac{\partial B_{r}}{\partial z}\right)-B_{\phi}\frac{\partial B_{\phi}}{\partial z}\right],
\end{multline}
\begin{equation}
\frac{1}{r}\frac{\partial}{\partial r}\left(rB_{r}\right)+\frac{\partial B_{z}}{\partial z}=0,
\end{equation}
\begin{equation}
\frac{\partial}{\partial z}\left[r\left(v_{z}B_{r}-v_{r}B_{z}\right)\right]=0,
\end{equation}
\begin{equation}
\frac{\partial}{\partial r}\left(v_{r}B_{\phi}-\frac{\lambda B_{r}}{r}\right)=\frac{\partial}{\partial z}\left(\frac{\lambda B_{z}}{r}-v_{z}B_{\phi}\right),
\end{equation}
\begin{equation}
\frac{\partial}{\partial r}\left[r\left(v_{z}B_{r}-v_{r}B_{z}\right)\right]=0,
\end{equation}
\begin{multline}
\frac{v_{r}}{\Gamma_{3}-1}\left[\frac{\partial p}{\partial r}-\Gamma_{1}\frac{p}{\rho}\frac{\partial \rho}{\partial r}\right]+\frac{v_{z}}{\Gamma_{3}-1}\left[\frac{\partial p}{\partial z}-\Gamma_{1}\frac{p}{\rho}\frac{\partial \rho}{\partial z}\right]\\=Q^{+}-Q^{-}=f_{m}Q^{+}. 
\label{eq:energy}
\end{multline}
Here, $F$ is the magnitude of gravitational force corresponding to the pseudo-Newtonian potential for a rotating black hole, $W_{ij}$ is the generalized viscous shearing stress which can be written using \cite{1973A&A....24..337S} prescription with appropriate modification given by \cite{1996ApJ...464..664C} and \cite{2003MNRAS.342..274M}, as $W_{r\phi}=\alpha (p+\rho v_{r}^{2})$; and the other components can be written following \cite{2009RAA.....9..157G} as $W_{\phi z}\approx \frac{z}{r}W_{r\phi}$ and $W_{rz}\approx \frac{z}{r}\alpha W_{r\phi}$. The first and second terms of the left-hand side of the equation~(\ref{eq:energy}) represent the radial and vertical advection of the flow respectively, where the details regarding adiabatic exponents $\Gamma_{1}$ and $\Gamma_{3}$ are given in \cite{2018MNRAS.476.2396M}.
The right-hand side of the equation~(\ref{eq:energy}) represents the difference between the net rates of energy generated ($Q^+$) and radiated out ($Q^-$) 
per unit volume, where the contribution in $Q^{+}$ comes from both viscous and magnetic parts as $Q^{+}=Q^{+}_{vis}+Q^{+}_{mag}$. The details regarding viscous contribution can be followed from the existing literature \citep[e.g.][]{1996ApJ...464..664C,2009RAA.....9..157G} and can be written as
\begin{equation}
Q^{+}_{vis}=\alpha \left(p+\rho v_{r}^{2}\right)\frac{1}{r}\left[\frac{\partial \lambda}{\partial r}-\frac{2\lambda}{r}+\frac{z}{r}\frac{\partial \lambda}{\partial z}+\alpha z\left(\frac{\partial v_{r}}{\partial z}+\frac{\partial v_{z}}{\partial r}\right)\right].
\end{equation}
The annihilation of the magnetic fields and an abundant supply of magnetic energy are responsible for magnetic heating and can be written as \citep[e.g.][]{1974Ap&SS..28...45B,1998RvMP...70....1B}
\begin{multline}
Q^{+}_{mag}=\frac{1}{4\pi}\left[B_{r}B_{z}\left(\frac{\partial v_{r}}{\partial z}+\frac{\partial v_{z}}{\partial r}\right)+B_{\phi}B_{r}\left(\frac{1}{r}\frac{\partial \lambda}{\partial r}-\frac{2\lambda}{r^{2}}\right)\right .\\ \left . +\frac{B_{\phi}B_{z}}{r}\frac{\partial \lambda}{\partial z}\right].
\end{multline}
The factor $f_{m}$ varies from $0$ to $1$ \citep{2010MNRAS.402..961R},
indicating the degree to which the flow is cooling-dominated or advection-dominated respectively. For the present purpose, we hypothesize $f_m$ to be
$0.5$ \citep{2015ApJ...807...43M}. In this coupled disc-outflow model, we also make a reasonable hypothesis that within the disc flow region, the vertical variation of any dynamical variables is much less than that with radial variation, which allows us to choose $\partial A/\partial z \approx sA/z$, for any dynamical variable $A$ and $s$ is just degree of scaling for that corresponding variable.



\section{ENERGETICS OF THE ACCRETION INDUCED OUTFLOWS}

In \cite{2018MNRAS.476.2396M}, we emphasized that the presence of large scale strong magnetic field could change the disc flow behaviours drastically in the
 vicinity of black hole event horizon. Here, as an immediate observational consequence, we address the most efficient way to facilitate the outflow formation from the disc threaded by large scale strong magnetic fields. This magnetically
dominated outflow prone disc could easily explain the power observed in 
ULXs, without invoking IMBHs or super-Eddington accretions. Theoretically, the problem regarding magnetic field generation is still ill-understood. It was suggested that externally generated magnetic field can be captured and dragged inward through continuous accretion process \cite[e.g.][]{1974Ap&SS..28...45B,1976Ap&SS..42..401B}. The field becomes dynamically dominant in the vicinity of black hole through flux freezing due to inward advection of magnetic flux. Indeed, there is an upper limit to the amount of magnetic flux what the disc around a black hole can sustain. Comparing the energy density of the magnetic field with that of the accreting plasma giving rise to the corresponding Eddington luminosity near vicinity of a black hole provides an upper limit to the magnetic field and is known as Eddington magnetic field \citep{2010PhyU...53.1199B}, which can be expressed as $B_{Edd}\approx 10^{4} \ \text{G} \ \left(\frac{M}{10^{9}M_{\odot}}\right)^{-1/2}$. 

\begin{figure}
	\includegraphics[width=\columnwidth]{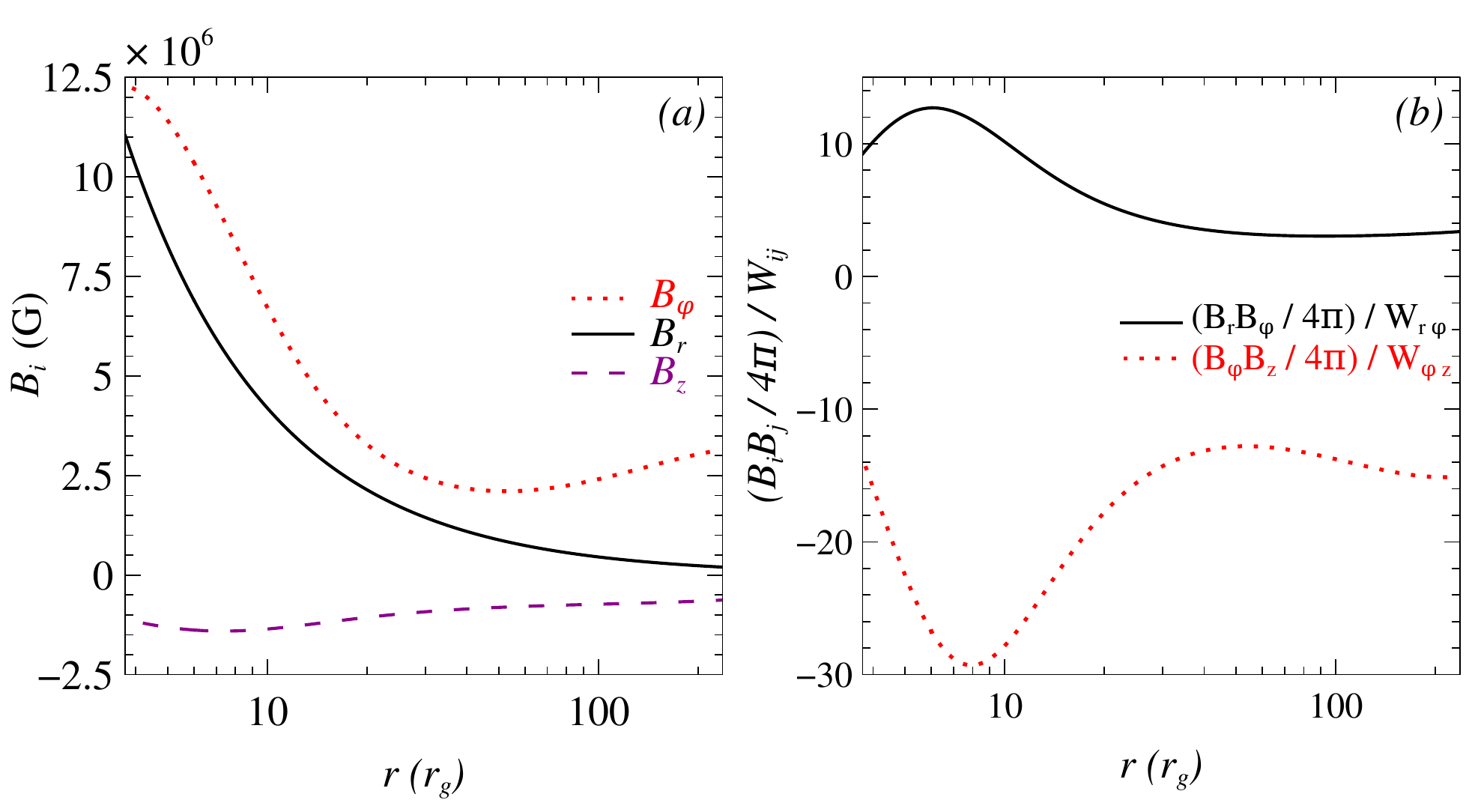}\caption{The variation of $(a)$ magnetic field components, and $(b)$ magnetic- to viscous-stress ratios, as functions of radial coordinate. The other parameters are $M_{BH}=20 \ M_{\odot}, \ \dot{M}=0.05 \ \dot{M}_{Edd}$, $f_{m}=0.5$ and $\alpha=0.015$. Note that 
$\left(B_r B_z/4\pi\right)/W_{rz}\sim 1000$ and, hence, is not shown here.} 
	\label{fig:field}
\end{figure}

By solving the set of model equations described in \S 3, we obtain various 
flow characteristics. The detailed solutions, in the spirit of its
$1.5-$dimensional counterpart \citep{2018MNRAS.476.2396M}, will be presented in a future work (Mondal \& Mukhopadhyay, in preparation).
For the present purpose, we concentrate on a few special physics of it.
First, the different field components are shown in Fig.~\ref{fig:field}a for a typical case. Here we choose the field profiles and their maximum possible magnitudes at the critical point in order to sustain at least one inner saddle type critical point \cite[see][for details]{2018MNRAS.476.2396M}. The maximum attainable field strength here is of the order of 
few factor times $10^{7}$ G for a $20 \ M_{\odot}$ non-rotating black hole. Note interestingly that advection of both poloidal and toroidal magnetic fields is happening, unlike pure MAD \citep{2003PASJ...55L..69N}.

Another important criterion to decide whether the accreting gas is gravitationally bound to the black hole or not, could be the Bernoulli parameter, $b$, of the gas \citep[e.g.][]{2012MNRAS.426.3241N}. This is the sum of kinetic energy, potential energy, gas enthalpy and contributions from viscous and magnetic shear 
stresses, and, can be defined as
\begin{multline}
	b= \frac{v^{2}}{2}+\frac{\gamma}{\gamma -1}     \frac{P}{\rho}+\Phi +\frac{1}{4\pi \rho} \left( B_{\phi}^{2}+B_{z}^{2}-\frac{v_{z}}{v_{r}} B_{r} B_{z} -\frac{\lambda}{rv_{r}} B_{r} B_{\phi} \right) \\
	-\frac{1}{\rho v_{r}} \left( \frac{\lambda}{r}W_{r\phi}+v_{z}W_{rz} \right).
\end{multline}
Here, $v^{2}=v_{r}^{2}+\lambda ^{2}/r^{2}+v_{z}^{2}$. The positive $b$ in this highly magnetized advective flows provides unbound matter and hence outflows.

In this disc-outflow symbiosis, the most important quantities we compute to study the energetics of the outflow, eventually contributing to observables
of ULXs, are the mass outflow rate and the power of the outflow extracted from the disc self-consistently in the presence of large scale strong magnetic field along with $\alpha$-viscosity.
The total mass accretion rate (sum of inflow and outflow rates) can be obtained by integrating the continuity equation along vertical and radial directions. Following \cite{2010NewA...15...83G}, the mass outflow rate can be written as
\begin{eqnarray}
\dot{M}_{j}(r)=-\int 4\pi r \rho (h_{surf}) v_{z} (h_{surf}) \  dr +c_{j},
\end{eqnarray}
where the constant $c_{j}$ is determined at the outer radius of the disc, outside which the outflow velocity is negligible $(v_{z}\simeq 0)$ and $h_{surf}$
is the disc-outflow coupled region's scale height. The quantity $\dot{M}_{j}(r)$ refers to the rate at which the outflowing mass flux ejects from the disc-outflow surface ($h_{surf}$). We restrict our model and calculations vertically up to the disc-outflow surface region, above which the outflow becomes decoupled and accelerates. The outflow power extracted from the disc is defined as the combination of mechanical, enthalpy, viscous and the Poynting parts and can be expressed as
\begin{multline}
	P_{j}(r)=\int 4\pi r  \Bigg[ \rho v_{z} \left\lbrace  \frac{v^{2}}{2}+\frac{\gamma}{\gamma -1}     \frac{P}{\rho}+\Phi - \Bigg(\frac{\lambda}{r}W_{\phi z}+v_{r}W_{rz}\Bigg) \right\rbrace  \\
	+\frac{v_{z}}{4\pi} \Bigg( B_{r}^{2}+B_{\phi}^{2}-\frac{v_{r}}{v_{z}} B_{r} B_{z} -\frac{\lambda}{rv_{z}} B_{\phi} B_{z} \Bigg) \Bigg]_{h_{surf}} dr.
\end{multline}

\begin{figure}
	\includegraphics[width=\columnwidth]{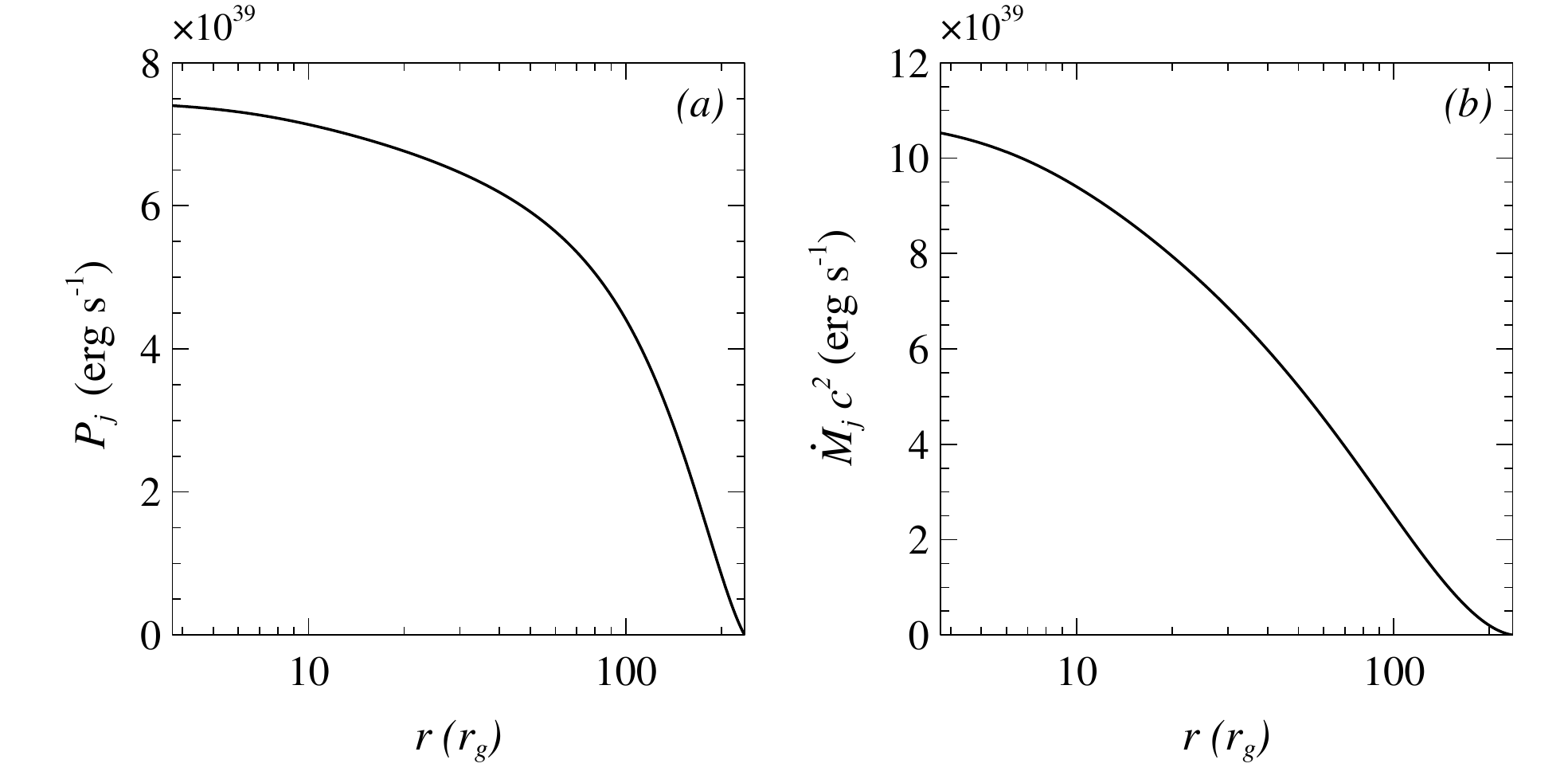}\caption{The variation of $(a)$ the outflow power, and $(b)$ the rate of rest mass energy associated with outflow, as functions of radial coordinate. The other parameters are same as in Fig.~\ref{fig:field}.} 
	\label{fig:power}
\end{figure}

The variations of this outflow power and the rate of rest mass energy associated with outflow are shown in Fig.~\ref{fig:power}. It indicates the outflow power at an arbitrary $r$ obtained by integrating from outer radius to that $r$. However, the observed power is expected to be liberated from inner region only. To compute these, we consider a stellar mass black hole of mass $M_{BH}=20 M_{\odot}$ with total mass accretion rate at infinity, $\dot{M}=0.05  \dot{M}_{Edd}$, where $\dot{M}_{Edd}=L_{Edd}/\eta c^{2}=1.39\times 10^{18} (M_{BH}/M_{\odot}) \ g \ s^{-1}$, considering radiative efficiency $\eta=0.1$ for a non-rotating black hole. Fig.~\ref{fig:power}a indicates that the outflow power at the outer region of the disc, far away from the black hole, is very small and it increases monotonically towards the central region. This is due to the negligible outflow velocity at the outer disc region, beginning of sub-Keplerian accretion flow. On the other hand, the dynamically dominant magnetic field in the inner region enhances the outflow power. This magneto-centrifugally driven outflows from the disc threaded by the open magnetic field lines provide very efficient way to liberate the gravitational potential energy through accretion process. Also, the vertically inflating toroidal fields exert an outward pressure to provide a generic explanation of strong outflows. The maximum attainable power of this magnetically driven outflows is $\sim 7.5\times 10^{39}$ erg s$^{-1}$ for a non-rotating black hole. 
Very importantly, the contribution from magnetic stress is orders of 
magnitude larger than viscous stress, as shown in Fig.~\ref{fig:field}b. Hence, the 
outflow power and generally energetics are magnetically driven in practice, $\alpha-$viscosity hardly
plays any role in it. It can be safely assumed that a significant portion of this magnetically driven outflow power is reprocessed and converted to X-ray luminosities observed in ULXs. We plan to attempt for a more quantitative estimate of the conversion in the future work. The plan would be, e.g., to combine the present model with that of \cite{2010MNRAS.402..961R}.

\section{DISCUSSIONS}

In this $2.5$-dimensional geometrically thick, sub-Keplerian, magnetized, viscous, advective, coupled disc-outflow symbiosis, we address the role of large scale strong magnetic fields in order to explore the energetics of the accretion-induced outflow. 
These energetics are essentially expected to constrain the ULX power. 
Accretion  discs can carry small as well as large scale magnetic fields. The strength of magnetic field plays indispensable roles in the dynamics of the accretion and hence outflow parameters. Now a question automatically arises: is there any upper limit to the amount of magnetic flux the disc around a black hole can sustain? In the accretion environment, the generation of seed magnetic field from zero field condition and its enhancement is very common through some well known existing mechanisms (e.g. Biermann battery mechanism, differential rotation, turbulence, dynamo process etc). However, the large scale magnetic field generally can not be produced in the disc. Nevertheless, it was suggested that the externally generated field can be captured from environment, say, companion star or interstellar medium, and dragged inward by the accreting plasma \citep[e.g.][]{1974Ap&SS..28...45B,1976Ap&SS..42..401B}. This magnetic field is greatly compressed and becomes dynamically dominant through flux freezing due to continued inward advection of the magnetic flux in this quasi-spherical accretion flow. The beauty of this idea is that we do not have to worry about the upper limit of the strength of the magnetic field threading the disc. It automatically sets up through magneto sonic/critical point analysis \citep[see][]{2018MNRAS.476.2396M} at the critical point and then evolves self-consistently. In this computation, we find the magnetic field strength near inner most region of the accretion flow is of the order of a few
factor times $\sim 10^{7}$ G for stellar mass black holes. However, this field strength is quite below the Eddington magnetic field limit $B_{Edd}\simeq 7.07\times 10^{7}$ G, for $20 \ M_{\odot}$ black holes. Based on several observational and theoretical modelling, the typical magnetic field strength in the black hole vicinity is of the order of $B\approx 10^{8}$ G for stellar mass black hole and $B\approx 10^{4}$ G for supermassive black hole \citep[e.g.][]{2011AstBu..66..320P, 2016A&A...593A..47B}. Hence the required magnetic field strength in our scenario is perfectly viable. However, such magnetic fields are not always possible to capture either from companion stars or interstellar medium, thus explaining ULXs to be rare.

\section{CONCLUSIONS}

Actual source of energy in ULXs is still under debate. On the other hand,
the dynamics and energetics of the outflow of underlying systems are intrinsically coupled to the disc flow behaviors through the fundamental conservation laws (mass, momentum and energy). In this advective paradigm, the presence of large scale strong magnetic field provides a generic explanation of powerful
unbound matters. The unbounded matter in the form of outflow is more plausible to emerge from the hot, puffed up region of the accretion flow. Most of the energy released by accreting matter is available to drive an outflow.
The maximum possible outflow power in our model is $\sim 7.5\times 10^{39}$ erg s$^{-1}$ for a non-rotating stellar mass black holes accreting at sub-Eddington accretion flow. Hence, this scenario can give a plausible indication to visualise the unclear nature of hard-state ULXs without incorporating the existence of the missing class of intermediate mass black holes, nor with the super-Eddington accretion.


\section*{ACKNOWLEDGEMENTS}

BM acknowledges the hospitality of Max-Planck-Institute for Gravitational Physics,
Albert Einstein Institute, Potsdam-Golm, Germany, where part of the manuscript
was written.











\bsp	
\label{lastpage}
\end{document}